\documentclass[conference]{IEEEtran}
\usepackage[english]{babel}
\usepackage{enumerate}
\usepackage{makeidx}
\usepackage{url}
\usepackage{footnote}
\usepackage{subfigure}
\usepackage{algorithm}
\usepackage{algpseudocode}
\usepackage{latexsym}
\usepackage{amssymb}
\usepackage[dvips]{graphicx}
\usepackage{amsmath}
\usepackage{graphicx}
\usepackage{float}
\usepackage{comment}
\usepackage{graphicx}
\usepackage{graphicx}
\usepackage{cite}
\begin{document}
%
\title{Cloud Offloading for Multi-Radio Enabled Mobile Devices}
\author{\IEEEauthorblockN{S.~Eman~Mahmoodi}
\IEEEauthorblockA{Dept. of Electrical \\and Computer Engineering\\
Stevens Institute of Technology\\
Hoboken, New Jersey 07030--5991\\
Email: smahmood@stevens.edu}
\and
\IEEEauthorblockN{K.~P.~Subbalakshmi}
\IEEEauthorblockA{Dept. of Electrical \\and Computer Engineering\\
Stevens Institute of Technology\\
Hoboken, New Jersey 07030--5991\\
Email: ksubbala@stevens.edu}
\and
\IEEEauthorblockN{Vidya Sagar}
\IEEEauthorblockA{Dept. of Electrical \\and Computer Engineering\\
Stevens Institute of Technology\\
Hoboken, New Jersey 07030--5991\\
Email: vsagar@stevens.edu}}
\maketitle
\begin{abstract}
 The advent of 5G networking technologies has increased the expectations from mobile devices, in that, more sophisticated, computationally intense applications are expected to be delivered on the mobile device which are themselves getting smaller and sleeker. This predicates a need for offloading computationally intense parts of the applications to a resource strong cloud. Parallely, in the wireless networking world, the trend has shifted to multi-\emph{radio} (as opposed to multi-channel) enabled communications. In this paper, we provide a comprehensive \emph{computation} offloading solution that uses the multiple radio links available for associated data transfer, optimally. Our contributions include: a comprehensive model for the energy consumption from the perspective of the mobile device; the formulation of the joint optimization problem to minimize the energy consumed as well as allocating the associated data transfer optimally through the available radio links and an iterative algorithm that converges to a locally optimal solution. Simulations on an HTC phone, running a 14-component application and using the Amazon EC2 as the cloud, show that the solution obtained through the iterative algorithm consumes only 3\% more energy than the optimal solution (obtained via exhaustive search).
\end{abstract}
\IEEEpeerreviewmaketitle
\section{Introduction}
\IEEEPARstart{T}{he} ``anywhere, anytime" promise of 5G networking has created a large demand for more sophisticated applications on energy constrained mobile devices \cite{WuTalJohHimJoh11}, leading to a huge increase in computational demand on the end devices \cite{KumLu10}. Meanwhile, the promise of 5G networking has also seen a surge in mobile device generated web traffic.  In the year 2012 alone mobile web traffic increased by 70\% and is expected to grow up to 13 times by 2017. One solution to this problem is to offload computations to the more resource strong cloud infrastructure \cite{ValCro13,GuNahMes04,ChuIhmManNaiMayPat11}. 

The term cloud offloading can mean either data flow offloading in networking applications \cite{MerVaiZor08},\cite{BhaVai10} or offloading computation intense processes on to the cloud. In this paper, we refer to the latter. Cloud offloading can be classified into three categories: (a) those that always offload to the cloud \cite{etime}; (b) ``all or nothing offloading" where either the entire application is offloaded to the cloud or executed locally, typically using an energy threshold to decide between offloading and not \cite{stocccccc13,WenZha12}; and (c) piecewise decisions, where some parts are executed locally while the others are offloaded to the cloud \cite{HuaNiyjjjjjjjjjjj12,KovTiaKla12,mauiiiii10,KosAucHuiMorZha12}.
The third category offers the most flexibility for trade-offs, and can be done either at the coarse component level \cite{HuaNiyjjjjjjjjjjj12,WanVas12,XinwenSang11} or at finer, method \cite{mauiiiii10} or instruction levels \cite{BarSarLor13}.  
 
While computation offloading to a resource strong cloud seems like the natural solution to the resource crunch at the mobile device level, it is essential to take into account the associated data transfer that must take place between the components that are executed in the cloud and their counterparts in the mobile device. Given the already increasing demands on the wireless backbone caused by the promise of 5G networking, this means that \emph{computation offloading must be viewed in the context of the already increasing mobile traffic}. Hence it would be prudent to \emph{optimally} use \emph{all} of the radio interfaces (like  WiFi, 3G, HSPA, and LTE), as appropriate, that are available in the multi-radio equipped mobile devices of today. 

In this paper, we propose a solution that optimally decides which components of an application to offload and which to execute locally, \emph{while simultaneously optimizing the percentage of data (associated with this offloading) to be sent via each radio interface}. Given recent advances in technologies that enable bandwidth aggregation in wireless devices \cite{HonSenCha13},\cite{Kas12} our solution is implementable in practice. To the best of our knowledge, this is the first such solution that approaches cloud offloading for multi-radio enabled devices. Other works that fall under general umbrella of the radio-aware computation offloading include \cite{HuaNiyjjjjjjjjjjj12}, where the best of the available wireless interfaces is chosen (only one of the wireless interfaces) for data transfer, rather than a solution that considers using all of the radio interfaces simultaneously. In \cite{BarSarLor13} a cloud offloading scheduling mechanism is proposed for queue stability, but this work only deals with multi-channel systems, not multi-radio networks. Etime \cite{etime} is an ``everything on the cloud'' offloading strategy, which adapts to the condition of the wireless link, but this work does not consider multiple interfaces.

In this paper, we develop a comprehensive model for the energy consumed by the mobile device, including energy expended in communicating relevant data between the cloud and the device. We set up the computation offloading problem as a joint optimization to minimize the energy consumed on the device while at the same time maximizing the radio resources available to the device, under two constraints: (1) the total run time deadline of the application and (2) the maximum flow rate constraint on the radio resources. 
Since this optimization problem is non-linear and hence computationally intense, we also propose an iterative algorithm that converges to a local optimum. Simulations show that the proposed iterative algorithm performs very close to the optimal solution for a significant reduction in complexity.
\section{System Model}
Consider a mobile device with $K$ radio interfaces, running computationally intense applications with $M$ components (See Fig. 1, with an example where $K=2$ and $M=6$).
Any given component may require data from the other components to complete execution. This data dependency is determined based on the corresponding application call graph (dependency matrix). In this example, the optimal offloading strategy stipulates that Components 1, 2, 4 and 6 be executed in the mobile device, and Components 3 and 5 be offloaded to the cloud. In Fig. 1, Component 3 requires $d_{23}$ units of data from Component 2 to complete execution. In this example, 60\% ($\nu_{2,1}$ = 0.6) of this data is sent through the Radio Interface 1 (WiFi, say), and 40\% ($\nu_{2,2}=0.4$) through Interface 2 (LTE) to give us the most performance efficient offloading strategy. Once Component 3 and 5 have finished execution, the data needed by Component 6 from Component 5 ($d_{56}$) must be sent to the mobile device via one of the radio interfaces (for example in Fig. 1 is WiFi). We assume that only one radio interface is used for data reception ($\sum\limits_{k = 1}^K{\gamma_{i,k}}=1 $), leaving the optimization of radio resource allocation for the downlink as future work. Also, we assume that the energy consumption and the time required to transfer data within components that are executing in the same entity (whether cloud or mobile) is negligible in comparison to when the data must be transferred between entities. We also assume that the components of the application are executed in a predetermined manner \cite{HuaNiyjjjjjjjjjjj12},\cite{mauiiiii10}. This is not an unreasonable assumption as the compiler usually predetermines this order.

\begin{figure}[t]
	\centering
		\includegraphics[width=0.5\textwidth]{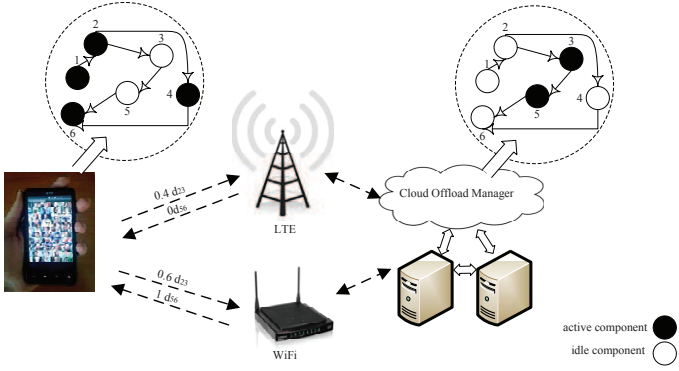} 
\caption{An example of application offloading to the cloud. In this figure, the dots represent components of the application. There are 6 components in this application. Components 1,2, 4, and 6 run on the device, whereas components 3 and 5 are executed in the cloud. Two radio links are available to the mobile device for offloading components to the cloud and the diagram shows the ratio of data that is sent via each radio interface. The terms active (idle) components refers to the components that are executed (or not) in that particular entity, mobile device or the cloud.}
\end{figure}

\begin{table}
\begin{savenotes}
\begin{center}
\caption{Parameter Definitions.}
    \begin{tabular}{ | l | p{5.5cm} |}
    \hline
    Parameters & Definitions \\ \hline
    $M$ & Number of components in the application. \\ \hline
		$K$ & Number of radio interfaces in the system model. \\ \hline
    $P_{\rm{ac}}^{\rm{(m)}}(i)$ & Power consumed by the mobile device when it is actively processing component $i$. \\ \hline
		$P_{{\rm{id}}}^{\rm{(m)}}$ & Power consumed by the mobile in the idle mode. \\ \hline
		 $P_k^{\rm{(Tx)}}$ $(P_k^{\rm{(Rx)}})$ & Transmit (Received) power consumed by the mobile device at radio interface $k$ . \\  \hline
    $\tau_i^{\rm{(m)}}$ $(\tau_i^{\rm{(c)}})$ & Time to process component $i$ in mobile (cloud).\\  \hline
    $\tau_{ij,k}^{\rm{(mc)}}$ $(\tau_{ij,k}^{\rm{(cm)}})$ & Time to transfer data required by component $j$ to mobile (cloud) from component $i$ in the cloud (mobile), using radio interface $k$.\\ \hline
		$T_i^{\rm{(com)}}$ & Time to transfer necessary data between the cloud and mobile, to execute component $i$. \\ \hline
$\alpha_{ij}$ & Component dependency indicator: 1 if component $i$ must be processed before j, 0 otherwise. \\ \hline
$I_i$ & Processing place indicator: 1 if component $i$ is processed on cloud, 0 if processed on mobile. \\ \hline
$\nu_{i,k}$ & Percentage of data upload using radio interface $k$, for execution of component $i$ in the cloud. \\ \hline
$\gamma_{i,k}$ & Radio receiving indicator: 1 if transferred data of component $i$ is received at radio $k$, 0 otherwise. \\ \hline
$d_{ij}$ & Data size required by component $j$ from $i$. \\ \hline
$R_k^{\rm{(d)}}$ ($R_k^{\rm{(u)}}$) & Downlink (Uplink) service rate for radio $k$. \\ \hline
$r_k$ & Demand rate for radio interface $k$. \\ \hline
$E_i$ & Total energy consumed by the mobile device to run component $i$. \\ \hline
$E_i^{\rm{(m)}}$ ($E_i^{\rm{(c)}}$) & Energy consumed by the mobile device to run component $i$ in the mobile (cloud). \\ \hline
$E_i^{\rm{(com)}}$ & Energy consumed by the mobile for data transfer of component $i$ between cloud and mobile. \\ \hline
    \end{tabular}
\end{center}
\end{savenotes}
\end{table}
The parameters needed to set up the optimization are described in Table I.
We model the energy consumed by the mobile device in running application component $i$, as $E_i= E_i^{\rm{(m)}}+E_i^{\rm{(c)}}+E_i^{\rm{(com)}}$,
where $E_i^{\rm{(m)}}$, $E_i^{\rm{(c)}}$ and $E_i^{\rm{(com)}}$ are all defined in Table I. The energy consumed to execute component $i$ locally, in the mobile device, is expressed as $E_i^{\rm{(m)}}=(1-I_i)P_{\rm{ac}}^{\rm{(m)}}(i)\tau_i^{\rm{(m)}}$. If the component is executed remotely, then the mobile will only spend the idle power for the duration of this execution. Hence, the energy consumed by the mobile when component $i$ is being remotely executed, is given by $E_i^{\rm{(c)}}=I_iP_{{\rm{id}}}\tau_i^{\rm{(c)}}$.
$E_i^{\rm{(com)}}$ comes into play when either the component immediately preceding the component $i$, or immediately succeeding component $i$ is executed in the other entity. $E_i^{\rm{(com)}}$ can be written as $E_i^{\rm{(com)}} = \sum\limits_{j=1}^M {\sum\limits_{k = 1}^K {(\alpha_{ij}\varepsilon_{ij,k}+\alpha_{ji}\varepsilon_{ji,k})}}$, where $\varepsilon_{ij,k}$  (or $\varepsilon_{ji,k}$) is the energy consumed in transferring data from component $i$ ($j$) to component $j$ ($i$) using radio interface $k$, when component $i$ ($j$) is executed immediately before component $j$ ($i$). They can be written as follows:
\begin{equation}\label{eqn:varEidc}
\begin{array}{l}
\varepsilon_{ij,k}=I_i(1-I_j)\gamma_{j,k}P_{{\rm{id}}}\tau_{ij,k}^{\rm{(cm)}}+(1-I_i)I_j\nu_{i,k}P_k^{\rm{(Tx)}}\tau_{ij,k}^{\rm{(mc)}},
\end{array}
\end{equation}
\begin{equation}\label{eqn:varEicd}
\begin{array}{l}
\varepsilon_{ji,k}=I_i(1-I_j)\nu_{j,k} P_{{\rm{id}}}\tau_{ji,k}^{\rm{(mc)}}+(1-I_i)I_j\gamma_{i,k}P_k^{\rm{(Rx)}}\tau_{ji,k}^{\rm{(cm)}}.
\end{array}
\end{equation}
The first terms on the RHS of equations \eqref{eqn:varEidc} and \eqref{eqn:varEicd} represent the idle powers consumed when the relevant component is being executed in the cloud, and second terms represent the energy consumed in transmitting or receiving the relevant data. The time needed to transfer data in the downlink communication (cloud to mobile) and uplink communication (mobile to cloud) are given by $\tau_{ij,k}^{\rm{(cm)}}=\frac{{d_{ij}}}{{R_k^{\rm{(d)}}}}$ and $\tau_{ij,k}^{\rm{(mc)}} = \frac{{d_{ji}}}{{R_k^{\rm{(u)}}}}$ respectively, where $R_k^{\rm{(d)}}$ and $R_k^{\rm{(u)}}$ are the downlink and uplink rates respectively, on radio interface $k$. $d_{ij}$ is the size of the data that must be transferred from component $i$ to $j$.

\section{Energy Efficient and Radio Resource Optimized Offloading}
\subsection{Problem Formulation}
In this section, we formulate an optimization problem to minimize the total energy consumed by the mobile user in executing a given application under total execution time constraints. Specifically, we will formulate an optimization problem that will determine which components should be executed where (in the device or cloud) and what percentage of data should be allocated to each radio link for necessary uplink data transfer. This minimization is subject to the following constraints: deadline on the execution time of the application; flow rate control on each radio link used for computation offloading; and the total value of data percentage allocated to the radio interfaces for each offloaded component. The optimization problem is mathematically formulated as
\begin{equation}\label{eqn:minE}
\mathop {{\rm{min}}}\limits_{\boldsymbol{\nu, I}} {\rm{ }}E\buildrel \Delta \over= {\sum\limits_{i = 1}^M {E_i } }, 
\end {equation}
\noindent where $\boldsymbol{I}=[I_1 I_2 ... I_M]$ and $\boldsymbol{\nu}$ is a matrix with entries $\nu_{i,k}$, $\forall i,k$ and  $I_i$'s and $\nu_{i,k}$'s are defined in Table I. The constraint on the total application execution time is given by 
\begin{equation}\label{eqn:consdeadline}
\sum\limits_{i= 1}^M {T_i}\le T_{\rm{req}},
\end {equation}
where $T_{\rm{req}}$ is the execution time deadline of the application, and $T_i=T_i^{\rm{(m)}}+T_i^{\rm{(c)}}+T_i^{\rm{(com)}}, \forall i$. $T_i^{\rm{(m)}}$ represents the time taken for component $i$ to execute in the mobile device, and is given by $T_i^{\rm{(m)}} = (1-I_i)\tau_i^{\rm{(m)}}$. Similarly, $T_i^{\rm{(c)}} = I_i\tau_i^{\rm{(c)}}$ is the time taken to execute component $i$ in the cloud. $T_i^{\rm{(com)}}$ is the time taken to complete the necessary data transfer for execution of component $i$, and is given by
\begin{align}\label{eqn:cloudexecution time}
\begin{split}
T_i^{\rm{(com)}}=\sum\limits_{j=1}^M{}\sum\limits_{k = 1}^K{}&I_i(1-I_j)(\alpha_{ji}\nu_{j,k}\tau_{ji,k}^{\rm{(mc)}}+\alpha_{ij}\gamma_{j,k}\tau_{ji,k}^{\rm{(cm)}})+ \\ 
&(1-I_i)I_j(\alpha_{ij}\nu_{i,k}\tau_{ij,k}^{\rm{(mc)}}+\alpha_{ji}\gamma_{i,k}\tau_{ij,k}^{\rm{(cm)}}).
\end{split}
\end{align}

\noindent This constraint allows us to take into consideration the potential time delays in sending and receiving the data related to each component via radio links ($T_i^{\rm{(com)}}, \forall i$) and trading it off optimally for energy consumption on the device.

In order for the system to be stable, the transmit data rate on the radio interfaces must be less than the service rate of each radio interface. This is represented by the second constraint:
\begin{equation}\label{eqn:consqos}
\sum\limits_{i = 1}^M {\sum\limits_{\scriptstyle j = 1 \hfill \atop
  \scriptstyle j \ne i \hfill}^M{\alpha_{ij}(1-I_i)I_j\nu_{i,k}r_k}< R_k^{\rm{(u)}}}, \forall k.
\end {equation}
The final constraint ensures that for each component, the total data allocations to the radio interfaces sums up to the total data that needs to be transferred, and is expressed as
\begin{equation}\label{eqn:consnu}
\sum\limits_{\scriptstyle j = 1 \hfill \atop
  \scriptstyle j \ne i \hfill}^M{\alpha_{ij} \sum\limits_{k = 1}^K {\nu _{i,k} \le 1}} ,\ \ \ \ \ \ \ \ \ \ \ \ \ \ \ \ \forall i.
\end {equation} 
\subsection{Proposed Solution}
The objective function of the optimization problem is represented in Eq. \eqref{eqn:minE} with the constraints in Eqns \eqref{eqn:consdeadline}, \eqref{eqn:consqos}, and \eqref{eqn:consnu}. The objective function and the constraint in Eq. \eqref{eqn:consqos} involve product terms of two non-negative variables, thereby forming a nonlinear convex function. Thus, the problem can be solved using MIP (Mixed Integer Programming) using Lagrangian multipliers: $\kappa, \zeta_k, \phi_i$, $\forall i,k$. The Lagrangian, $L=L(\boldsymbol{\nu, I}, \kappa, \boldsymbol{\zeta, \phi})$, is expressed as
\begin{equation}\label{eqn:lagrange}
\begin{array}{l}
L=\sum\limits_{i=1}^M{{E_i(\nu_{i,k}, I_i)}}+\kappa\sum\limits_{i=1}^M{(T_i(I_i, \nu_{i,k})-T_{\rm{req}})}+\\{ \;\; \; \;\; \;} \sum\limits_{k=1}^K{\zeta_k(\sum\limits_{i=1}^M{\sum\limits_{\scriptstyle j = 1 \hfill \atop \scriptstyle j \ne i \hfill}^M{\alpha_{ij}((1-I_i)I_j\nu_{i,k}r_k-R_k^{\rm{(u)}})}})}+\\{\; \; \;\; \;\;}\sum\limits_{i=1}^M{\phi_i(\sum\limits_{\scriptstyle j = 1 \hfill \atop \scriptstyle j \ne i \hfill}^M{\alpha_{ij}\sum\limits_{k=1}^K{\nu_{i,k}}}-1)}.
\end{array}
 \end{equation}

Minimizing $L$ will involve finding the best set of values for the parameters $\nu_{i,k}$, and $I_i$, $\forall i ,k$.
To obtain the best offloading policy (values of $I_i$), we write $L_i$ as a function of $I_i$ and a constant term ($c_1$) that does not depend on $I_i$. That is, $L_i=\Delta_i I_i+c_1$,
where 
\begin{equation}\label{eqn:delta}
\Delta_i=\Lambda_i+\sum\limits_{\scriptstyle j = 1 \hfill \atop
  \scriptstyle j \ne i \hfill}^M{(1-I_j)\Gamma_{i,j}^{\rm{(c)}}}-\sum\limits_{\scriptstyle j = 1 \hfill \atop
  \scriptstyle j \ne i \hfill}^M{I_j\Gamma_{i,j}^{\rm{(m)}}},
  \end{equation}
and $\Lambda_i$ is independent of $\nu_{i,k}$, and can be written as
\begin{equation}\label{eqn:Lambda}
\Lambda_i=P_{{\rm{id}}}\tau_i^{\rm{(c)}}-P_{\rm{ac}}^{\rm{(m)}}(i)\tau_i^{\rm{(m)}}+\kappa(\tau_i^{\rm{(c)}}-\tau_i^{\rm{(m)}}), 
\end{equation}
\noindent and
\begin{equation}\label{eqn:Gammac}
\Gamma_{i,j}^{\rm{(c)}}=(P_{{\rm{id}}}+\kappa)\sum\limits_{k=1}^K{(\alpha_{ji}\nu_{j,k}\tau_{ji,k}^{\rm{(mc)}}+\alpha_{ij}\gamma_{j,k}\tau_{ji,k}^{\rm{(cm)}})},
\end{equation}
and
\begin{equation}\label{eqn:Gammam}
\begin{array}{l}
\Gamma_{i,j}^{\rm{(m)}}=\sum\limits_{k=1}^K{}\big(\alpha_{ij}\nu_{i,k}P_k^{\rm{(Tx)}}\tau_{ij,k}^{\rm{(mc)}}+\alpha_{ji}\gamma_{i,k}P_k^{\rm{(Rx)}}\tau_{ij,k}^{\rm{(cm)}}+ \\
{ \;\; \;\;\; \;\;\; \;\;\; \; \;\; \;\;\; \;}\kappa(\alpha_{ij}\nu_{i,k}\tau_{ij,k}^{\rm{(mc)}}+\alpha_{ji}\gamma_{i,k}\tau_{ij,k}^{\rm{(cm)}})+\zeta_k\alpha_{ij}\nu_{i,k}r_k\big).
\end{array}.
\end{equation}
In Algorithm 1, we present an iterative algorithm to find the optimal values of $\nu_{i,k}$ and $I_i$ for each component. The algorithm is initialized with values for the Lagrange multipliers ($\kappa, \zeta_k, \phi_i$, $\forall i,k$) as well as an initial allocation of where the given component $i$ will be executed (values of $I_i$). The iteration index $r$ is set to 0, and the initial value of $I_i^{(r)}$ is given by:
\begin{equation}\label{eqn:lambda1}
I _i^{(r)}  = \left\{ {\begin{array}{*{20}c}
   1 & {\Lambda_i < 0},   \\
   0 & {\Lambda_i \ge 0}.  \\
\end{array}} \right.
\end{equation}
This initial schedule of components implies that the component $i$ will be scheduled to run in the cloud if the trade-off between energy consumption and execution time for running it on the cloud is favorable to running it on the mobile. 
To obtain optimum $\nu_{i,k}$s, we rewrite $L$ for Component $i$ and Radio Interface $k$ as: $L_{i,k}=\nu_{i,k}\Omega_{i,k}+c_2$, where $\Omega_{i,k}=\sum\limits_{j = 1}^M {\{\alpha_{ij}(1-I_i)I_j[\tau_{ij,k}^{\rm{(mc)}}(P_{k}^{\rm{(Tx)}}+\kappa)+\zeta_k r_k]+\phi_i\}},$
\noindent and $c_2$ is a constant w.r.t $\nu_{i,k}$. 
The optimal value of $\nu_{i,k}$, $\nu_{i,k}^{*}$ for a given value of $I_i$ is calculated as 
\begin{equation}\label{eqn:initial nu}
\nu_{i,k}^{*}  = \left\{ {\begin{array}{*{20}c}
   {(1-I_i)(1 - \frac{{\Omega _{i,k} }}{{\sum\limits_{i = 1}^M {\sum\limits_{k = 1}^K {\Omega_{i,k} } } }})} & {\sum\limits_{\scriptstyle j = 1 \hfill \atop
  \scriptstyle j \ne i \hfill}^M {\alpha _{ij} }  \ne 0}  \\
   0 & {\sum\limits_{\scriptstyle j = 1 \hfill \atop
  \scriptstyle j \ne i \hfill}^M {\alpha _{ij} }  = 0}  \\
\end{array}} \right.
\end{equation}
Now by using the value of $\nu_{i,k}^{*}$ by using \eqref{eqn:initial nu}, $I_i$ can be updated by
\begin{equation}\label{eqn:lambdak}
I_i^{(r)}=\left\{ {\begin{array}{*{20}c}
   1 & {\Delta_i < 0},  \\
   0 & {\Delta_i\ge 0}.  \\
\end{array}} \right.
\end{equation}

The iterations continue until Eq. \eqref{eqn:lagrange} is minimized. The algorithm converges, when the Langrange parameters have converged. The details are given in Algorithm 1.
\subsection{Convergance and Complexity of the Algorithm}
In line 1, $I_i$ and $\nu_{i,k}$ are initialized. In a nested loop, these two variable parameters are modified such that the Lagrangian formulation in Eq. \eqref{eqn:lagrange} is minimized. The strategy of Lines 3-17 of the algorithm has been discussed in subsection $B$. The variables $I_i$ and $\nu_{i,k}$ are opportunistically updated using Eqns \eqref{eqn:lambdak} and \eqref{eqn:initial nu}, respectively so that the objective function is minimized (lines 12,13 of the algorithm). The outer loop updates the Lagrangian multipliers using the subgradient method. Using the logic in \cite{BoyMut08}, we see that the updated multipliers ($\kappa$, $\zeta_{k}$, and $\phi_{i}$, $\forall i,k$) will converge to the optimum values of $I_i$ and $\nu_{i,k}$, $\forall i,k$.

Complexity of the modification loop (Lines 9-23) of the algorithm is O($r_{\rm{max}}M$), where $r_{\rm{max}}$ is the maximum number of iterations required to find the optimum vector $\boldsymbol{I}$. Note that we assume $M>K$. Overall, the complexity of the algorithm is O($s_{\rm{max}}r_{\rm{max}}M$), where $s_{\rm{max}}$ is the maximum required number of iterations to satisfy all the constraints in the optimization problem. The value of $s_{\rm{max}}$ depends on the initial values in line 6 and $\epsilon$ values in lines 28, 29 of the algorithm. In the simulations (Section VI), we observe that the mean values of $s_{\rm{max}}$ and $r_{\rm{max}}$ are 3 and 2, respectively. The complexity of the exhaustive search method is O($2^M\times k$), which is prohibitively high. 
\begin{algorithm}
\caption{Proposed Radio Aware Offloading Schedule.}
\begin{algorithmic} [1]
\State {\bf{initialization}}:
  \State\hspace{\algorithmicindent}      Set $r \gets 0$, modification index, $s \gets 1$ 
  \State\hspace{\algorithmicindent}    Set $I_i^{(0)}$ using Eq. \eqref{eqn:lambda1} 
  \State\hspace{\algorithmicindent}    Set $\Delta{^{(0)}}$ using Eq. \eqref{eqn:Lambda} 
  \State\hspace{\algorithmicindent}  Set $\nu_{i,k}^{(0)}$ using Eq. \eqref{eqn:initial nu}
  \State\hspace{\algorithmicindent}  Set initial values for parameters $\kappa ^{(s)}, \zeta_k^{(s)}, \phi _i^{(s)}$ 
  \State\hspace{\algorithmicindent}     Set $X_r= X_s \gets$False 
\Repeat {:}
\If{$\Delta_i^{(r)}<0$, $\forall i$}
\While{$X_r$=False}
\State calculate $\Delta_{i}^{(r+1)}=\Delta_{i}|_{I_i =I_i^{(r)},\nu_{i,k}=\nu_{i,k}^{(r)}}$ by \eqref{eqn:delta}
\State calculate $I_{i}^{(r+1)}$ by Eq. \eqref{eqn:lambdak}
\State calculate $\nu_{i,k}^{(r+1)}$ by Eq. \eqref{eqn:initial nu}
\If {$\exists i:\Delta_i^{(r+1)}\Delta_i^{(r)}<0$}
\State $\bf{Find}$ $\mathop {{\rm{min}}}\limits_{\tilde i} (\Delta _i^{(r+1)}; \forall i)$
\State $I_{\tilde i} \to 1 - I_{\tilde i}$,
\EndIf
\If {$\sum\limits_{i=1}^M L_i^{(r+1)}\ge\sum\limits_{i=1}^M L_i^{(r)}$}
\State $X_r$=True,
\EndIf
\State $r \to r + 1$,
\EndWhile
\EndIf
\State $\kappa ^{(s + 1)}  = \kappa ^{(s)}  - \varepsilon _\kappa  (T_{\rm{req}}  - \sum\limits_{i = 1}^M {T_i } )$
\State {$\zeta_k^{(s + 1)}=\zeta_k^{(s)}-\varepsilon_\zeta\times$\\ \hspace{18 mm}$(R_k^{\rm{(u)}}-\sum\limits_{i=1}^M\sum\limits_{\scriptstyle j = 1 \hfill \atop
\scriptstyle j\ne i \hfill}^M{|I_i-I_j|(1-I_i)\alpha_{ij}\nu_{i,k} r_k }), \forall k$}
\State $\phi _i^{(s + 1)}  = \phi _i^{(s)}  - \varepsilon _\phi  (1 - \sum\limits_{\scriptstyle j = 1 \hfill \atop
  \scriptstyle j \ne i \hfill}^M {\alpha _{ij} \sum\limits_{k = 1}^K {\nu _{i,k} } } ){\rm{  }}, \forall i$
\If{$\frac{{|\kappa ^{(s + 1)}- \kappa ^{(s)}|}}{{\kappa ^{(s + 1)}}}<\varepsilon_\kappa$ \& $\frac{{|\zeta _k^{(s + 1)}-\zeta_k^{(s)}|}}{{\zeta_k^{(s + 1)}}}<\varepsilon_\zeta$ \& \\ \hspace{9 mm}$\frac{{|\phi _i^{(s + 1)}- \phi _i^{(s)}|}}{{\phi_i^{(s + 1)}}} < \varepsilon _\phi,  \forall i, k$} 
\State $X_s$ = True,
\EndIf
\State $s \to s + 1$
\Until{any constraint in Eqs \eqref{eqn:consdeadline},\eqref{eqn:consqos},\eqref{eqn:consnu} is not satisfied: ($X_s$=False).}
\end{algorithmic}
\end{algorithm}
\section{Performance Analysis}
In this section, we investigate the efficiency of the proposed approach using an HTC Vivid smartphone with a 1.2 GHz dual core processor. This phone is equipped with two radio interfaces (k = 2): WiFi, and LTE. Moreover, we assume that whereas LTE is always available, the WiFi interface can sometimes be unavailable (as is common in real life scenarios). A multi-component video navigation application was used for the experiments. This application uses video processing, face detection, graphics, and clustering the video points. Graphics library tools are used from the  OpenGL mobile Android applications \cite{Open14}, face detection is used from \cite{AndFac14}, and all of the video processing features are available in \cite{OCV14}. We used fourteen component applications to form the codeset in our work. Note that the first and last components are executed in the mobile device, because most mobile initiated applications must start in the mobile device and usually have an output/display that happens on the mobile device. We measured execution time of the components in the HTC phone and the cloud, uplink and downlink rates and delays for WiFi and LTE. We obtained the dependency matrix of this application, and the size of the data that needs to be transferred between components. The Amazon Elastic Compute Cloud (Amazon EC2) was used for cloud computing capacity \cite{amazon13}. 
The average transmit power levels of the mobile device for WiFi, and LTE services are 300 and 600 mWs, respectively. The average received power levels were 100 and 250 mWs, respectively. The active and idle power levels of the phone are 644.9 and 22 mWs, respectively. The power consumption for the last component in the mobile device was 55 mWs. These power measurements are obtained by using “CurrentWidget: Battery monitor” application \cite{currentwidget}. The average wireless service rates for WiFi, LTE are 0.80 and 2.96 Mbps for the uplink transmission and 1.76 and 4 Mbps for the downlink transmission, respectively. Also, local execution time of the fourteen components are measured as [30 340 345 125 30 80 70 30 185 125 650 571 904 56] ms. The number of arriving requests is modeled as a Poisson distributed variable with average rate of 1.5 Mbps. The initial multiplier values for $\kappa, \phi$ and $\zeta$ were set to 0.1, 0.1, and $10^{-6}$, respectively. The results shown are averages of 1000 independent test runs.

Four scenarios are compared in this section. First, we consider the scenario that all components are executed locally in the mobile. The energy consumed in this scenario is used to normalize all energy values. The second scenario consists of executing the entire application on the cloud (other than the first and the last components). In this scenario, all data must be uploaded to the cloud. The third scenario is a brute force exhaustive search for the best values of $I_i$ for each component. That is, we manually schedule components $i =2$ through $13$ to run on either the cloud or the mobile and calculate the associated energy and time. Note, that since the first and last component must run on the mobile, we are left with $2^{(14-2)}$ combinations of possible values for the $I_i$'s. For each combination of $\bf{I}$, the problem turns out to be a linear optimization over the variable set $\boldsymbol{\nu}$. Thus, the radio allocation percentages are calculated using linear programming. The sets of $I_i$ and $\nu_{i,k}$, $\forall i, k$, values which minimize the energy consumption give the over all optimal solution. The approach in this scenario is called ``Exhaustive search". Finally, the fourth set of results is obtained by our iterative algorithm.

\begin{figure}[t]
	\centering
		\includegraphics[width=0.5\textwidth , height=0.22\textheight]{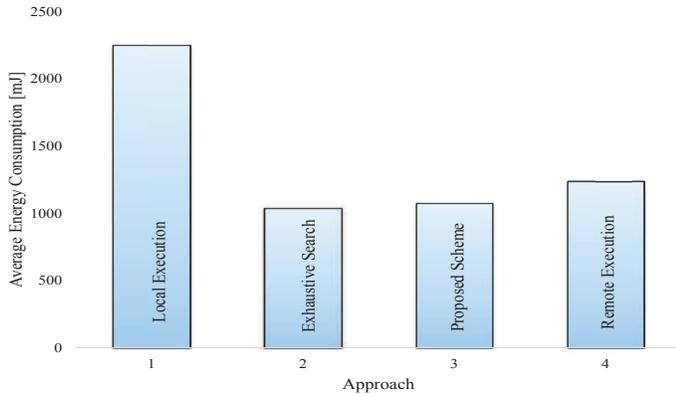}
	\caption{\footnotesize{Average Energy Consumption of the Four Approaches, while execution time equals 3.54 seconds.}}
\label{fig:energy}
\end{figure}

\begin{figure}[t]
	\centering
		\includegraphics[width=0.5\textwidth , height=0.22\textheight]{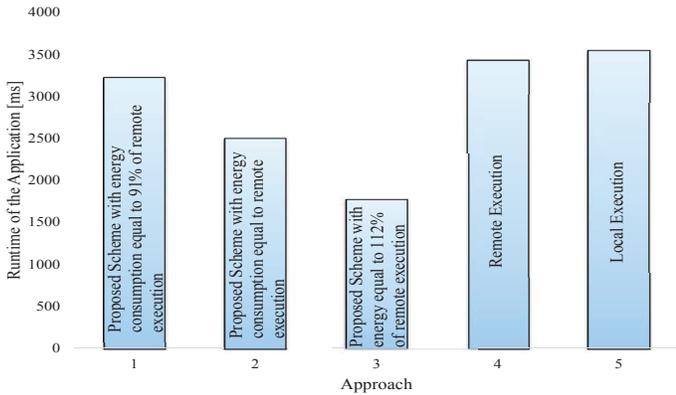}
	\caption{\footnotesize{Execution time of Different Approaches. }
}
\label{fig:execution time}
\end{figure}

\begin{figure}[t]
	\centering
		\includegraphics[width=0.5\textwidth]{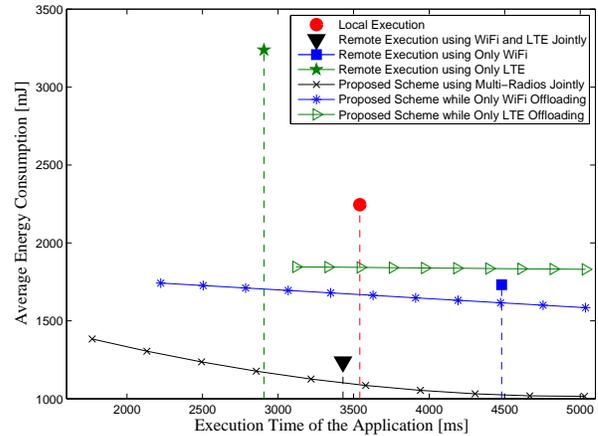}
	\caption{\footnotesize{Average Energy Consumption versus execution time of the application. Jointly, it shows the trade-off between the cost of energy consumption and execution time in the proposed scheme. We observe that the application can be executed in half of the time (0.52) it takes to be executed in the cloud with the cost of 12\% more energy consumption, and also the application can be executed with 20\% more energy saving in comparison to the remote execution with the cost of 42\% execution time extension in comparison to remote execution.}}
\label{fig:energy-execution time}
\end{figure}

Fig.~\ref{fig:energy} shows the average energy consumption for four different approaches while the application execution time  equals to 3.54 seconds. We observe that the proposed approaches (exhaustive search and the proposed iterative algorithm) result in lower energy consumptions in comparison to the others. Note that 3.54 seconds is the minimum execution time to execute the application locally, so that the execution time deadline is satisfied in all of the approaches. On an average, the proposed iterative algorithm consumes 3$\%$ more energy in comparison to the proposed optimal solution (Exhaustive search approach) for $T_{\rm{req}}=580$ ms. This is a fairly good trade-off for the reduced complexity of the proposed iterative algorithm. 
Fig.~\ref{fig:execution time} presents the execution time of different approaches in different scenarios. While local and remote execution approaches require longer application execution time, the proposed scheme gives us 29\% and 27\% faster execution time in comparison to these approaches respectively with the same amount of energy consumption to the remote execution approach. If we desire to save 9\% of energy, then we have still 9\% and 6\% faster execution time in comparison to local and remote execution respectively. On the other hand, if only fast execution of the application is important for us, then by costing 11\% more energy than remote execution, we can achieve 50\% and 48\% faster run rather than local and remote execution.
Fig.~\ref{fig:energy-execution time} plots the energy-execution time trade-off in the proposed scheme in comparison to the local and remote execution, while the proposed scheme takes advantage of three scenarios for radio resources: 1. WiFi and LTE are used jointly; 2. only WiFi is used for offloading; and 3. only LTE is used for offloading. The four points in the plot show local and remote executions by using only LTE, only WiFi, or both. We see that although remote execution by using LTE consumes much more energy in comparison to the others, the execution time for this scenario would be less than the others. Thus, there is a trade off between energy consumption and execution time of the application which is relied on the delay of offloading. On the other hand by using the proposed offloading scheme lesser energy is consumed with reasonable value for execution time. When the execution time deadlines are longer, there is more flexibility in offloading jobs to the cloud and hence energy consumptions for the mobile device reduces. Also, it is clear that joint use of radio resources gives less energy consumption and requires less execution time.

\begin{figure}[t]
	\centering
		\includegraphics[width=0.5\textwidth]{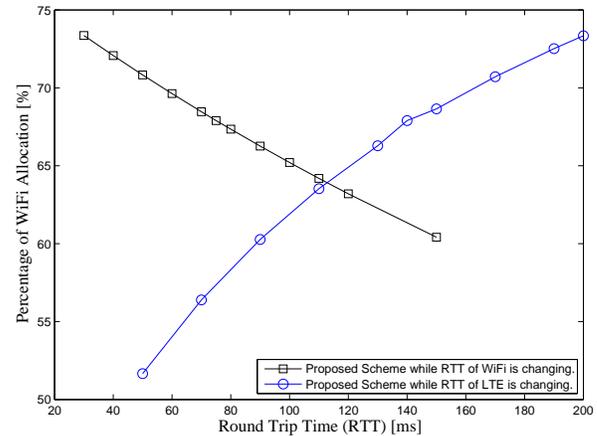}
	\caption{\footnotesize{Percentage of WiFi Allocation in the Proposed Scheme versus Round Trip Time (RTT) of WiFi and LTE.}}
\label{fig:percentage}
\end{figure}

Fig.~\ref{fig:percentage} plots the percentage of data stream to the cloud through WiFi (radio interface 1) versus RTT of the WiFi and LTE in the proposed scheme. We observe that by increase of RTT in WiFi for the range of 40-160 ms, less data stream is allocated to WiFi and more data stream is allocated through LTE for computation offloading. On the other hand, when RTT of LTE increases in the range of 50-200 ms, more data stream is allocated to WiFi and less data is allocated to LTE. Finally, 
\section{Conclusion}
We studied the problem of offloading computationally intense applications from mobile devices to a cloud infrastructure for multi-radio equipped mobile devices. We presented a comprehensive model for the energy consumed in offloading components to  the cloud. We modeled the decision to offload any given component to the cloud as an optimization problem that seeks to resolve the conflicting goals of reducing computation costs while keeping the execution time of the application below its deadline. We showed that this is a non-linear optimization problem. We proposed an iterative algorithm to find the local optima for the offload schedule of the components as well as the percentage of the data to be carried on each radio interface. We showed that the proposed algorithm consumes within 4\% of the optimal solution (obtained via brute force search) and also offers 31\% less energy consumption in comparison to offloading the entire application to the cloud.  
\ifCLASSOPTIONcaptionsoff
  \newpage
\fi

\bibliographystyle{IEEEtran}

\bibliography{cloud6}

\end{document}